\documentclass[10pt]{article}
\usepackage{amsmath,amssymb, graphicx}
\usepackage{bm, cite}
\usepackage{booktabs}   
\usepackage{siunitx}    
\usepackage{array}
\usepackage{url}
\usepackage{listings}
\usepackage[margin=1in]{geometry}

\title{And Yet Another FEM-Based Mode Solver for Dielectric Waveguides}
\author{Ergun Simsek \\
Department of Computer Science and Electrical Engineering \\
University of Maryland Baltimore County, Baltimore, MD 21250 \\
simsek@umbc.edu}
\date{\today}

\begin{document}
\maketitle
\begin{abstract}
We present a full-vector finite element method (FEM) mode solver for dielectric waveguides based on a mixed Nédélec--Lagrange discretization of Maxwell’s curl equations in the frequency domain. The formulation combines edge elements for transverse field components with nodal elements for the longitudinal component, enabling accurate modeling of hybrid modes while effectively suppressing spurious solutions. The solver is implemented in both MATLAB and Python with an emphasis on reproducibility, computational efficiency, and accessibility, including compatibility with cloud-based platforms. Numerical validation is performed on representative waveguide structures, demonstrating excellent agreement with COMSOL Multiphysics, with relative errors below $0.05\%$. Convergence studies confirm the expected accuracy trends with mesh refinement, while highlighting the trade-off between computational cost and precision. The proposed implementation provides a flexible and reliable open-source tool for integrated photonics research and education.
\end{abstract}
\section{Introduction}
The modal analysis of dielectric optical waveguides, which is crucial for integrated photonic device design, has been extensively studied using a wide variety of numerical techniques, among which the finite element method (FEM) has emerged as one of the most powerful and flexible approaches due to its geometric flexibility and ability to handle complex material distributions. 

Finite element formulations for optical waveguides can generally be derived either directly from Maxwell’s differential equations \cite{Koshiba92, Lee94, SelleriPetracek2001, Selleri2001, Simsek2025}  or through weighted residual \cite{Rahman2002} and variational principles \cite{Wu1985, VD2003}. Among these, the variational approach is particularly advantageous because it naturally leads to symmetric eigenvalue problems and provides a physically meaningful functional whose extremization corresponds to the propagation constant or frequency of the guided modes. Weighted residual methods, on the other hand, are more general and can be applied even when a variational formulation is not readily available.

Scalar FEM formulations, typically based on longitudinal field components such as $E_z$ or $H_z$, have historically been used for weakly guiding structures and quasi-TE or quasi-TM modes \cite{Koshiba92}. While computationally efficient, scalar approaches are fundamentally limited in their ability to represent hybrid modes in strongly guiding, anisotropic, or highly inhomogeneous structures. Consequently, their applicability is restricted in modern integrated photonics where high index contrast waveguides dominate. To overcome these limitations, full-vector finite element formulations based on Maxwell’s curl equations have been developed \cite{Lee94, Selleri2001, SelleriPetracek2001, Murphy2008}. These methods solve simultaneously for all three components of the electromagnetic field and are capable of accurately modeling hybrid modes in arbitrary dielectric waveguides. A widely used approach is the full-vector magnetic field ($\mathbf{H}$) formulation, which leads to a real symmetric eigenvalue problem under lossless conditions. However, a major challenge associated with node-based vector FEM is the emergence of non-physical or spurious modes that contaminate the numerical spectrum.

Several strategies have been proposed to eliminate spurious solutions. One classical approach is the penalty function method \cite{koshiba1987}, which enforces the divergence-free condition of the magnetic field in a weak sense by introducing a penalty term into the variational formulation. While effective in suppressing spurious modes in lossless cases, this approach is not always suitable for lossy or complex-valued eigenvalue problems.
An alternative formulation based on transverse field components was proposed using Galerkin methods, which successfully reduces spurious modes in anisotropic waveguides \cite{Angkaew1987}. However, this approach leads to a significant increase in computational complexity due to denser matrices and, in many cases, results in quadratic eigenvalue problems. These drawbacks limit its practical application in large-scale photonic simulations.
A major advancement in FEM waveguide modeling was the introduction of edge elements, which ensure tangential field continuity and inherently satisfy the divergence-free condition. Edge-based formulations significantly reduce spurious solutions and have become a standard in computational electromagnetics \cite{Nuno1997}. Hybrid formulations combining edge elements for transverse fields and nodal elements for longitudinal components further improve accuracy and allow for the consistent modeling of true hybrid modes in complex waveguide structures. Despite their advantages, edge element methods typically require higher computational cost and lead to more complex eigenvalue systems.
Later developments in computational electromagnetics have focused on improving robustness and accessibility of full-vector FEM solvers for integrated photonics applications. In particular, modern implementations aim to balance numerical accuracy with computational efficiency, enabling simulation of realistic waveguide geometries such as trapezoidal or irregular cross-sections that arise in fabrication processes. In this context, finite-difference-based solvers, while efficient for rectangular geometries \cite{Simsek2025}, are often insufficient for accurately resolving complex boundaries, motivating the continued use and development of FEM-based eigenmode solvers.

In this work, we build upon established full-vector FEM formulations for frequency-domain Maxwell’s equations under the harmonic assumptions $e^{j\omega t}$ and longitudinal propagation $e^{-j\beta z}$. Although the formulation itself is classical and well documented in the literature, including works such as \cite{Wu1985, Koshiba92, Lee94, VD2003, SelleriPetracek2001, Simsek2025}, the objective here is not to introduce a new theoretical contribution. Instead, the focus is on developing an efficient, reproducible, and open numerical implementation suitable for MATLAB and Python environments, including cloud-based platforms such as Google Colab. 

\section{Numerical Formulation and Implementation Details}
We consider a two-dimensional configuration by assuming that the waveguide extends infinitely along the \( z \)-axis. We start from Maxwell's curl equations in a source-free, linear, isotropic medium,
\[
\nabla \times \mathbf{E} = -j\omega \mu_0 \mu_r \mathbf{H},
\qquad
\nabla \times \mathbf{H} = j\omega \epsilon_0 \epsilon_r \mathbf{E},
\]
one eliminates $\mathbf{H}$ to obtain the vector wave equation
\[
\nabla \times \left( \mu_r^{-1} \nabla \times \mathbf{E} \right) - k_0^2 \epsilon_r \mathbf{E} = \mathbf{0},
\]
where $k_0 = \omega/c_0$ is the free-space wavenumber and $c_0$ is the speed of light in vacuum. For a waveguide structure invariant along $z$, every field component carries the factor $e^{-j\beta z}$, where $\beta$ is the propagation constant to be determined. The electric field is accordingly decomposed as
\[
\mathbf{E}(x,y,z) = \bigl[\mathbf{E}_t(x,y) + \hat{z}\,E_z(x,y)\bigr]\,e^{-j\beta z},
\]
where $\mathbf{E}_t = E_x\hat{x} + E_y\hat{y}$ is the transverse part. Under this decomposition, the three-dimensional curl splits into transverse and longitudinal contributions. For a vector
field $\mathbf{F}_t + \hat{z}F_z$ multiplied by $e^{-j\beta z}$ one has
\[
\nabla \times \mathbf{F}
= \left(\nabla_t \times \mathbf{F}_t\right)\hat{z}
  + \nabla_t F_z \times \hat{z}
  - j\beta\,\hat{z} \times \mathbf{F}_t,
\]
where $\nabla_t = \hat{x}\partial_x + \hat{y}\partial_y$ and
$\nabla_t \times \mathbf{F}_t = \partial_x F_y - \partial_y F_x$ denotes the scalar (out-of-plane) curl. Applying this decomposition twice inside the wave equation and separating transverse and longitudinal parts yields the coupled system
\begin{align}
\frac{1}{\mu_r}\left[
  \nabla_t\!\left(\nabla_t \cdot \mathbf{E}_t\right)
  - \nabla_t^2 \mathbf{E}_t
  - j\beta\,\nabla_t E_z
\right] + j\beta\,\frac{1}{\mu_r}\nabla_t E_z
- \frac{\beta^2}{\mu_r}\mathbf{E}_t - k_0^2\epsilon_r \mathbf{E}_t &= \mathbf{0},
\label{eq:transverse}\\
\frac{1}{\mu_r}\left[
  -\nabla_t^2 E_z + j\beta\,\nabla_t \cdot \mathbf{E}_t
\right] - k_0^2\epsilon_r E_z &= 0.
\label{eq:longitudinal}
\end{align}
Rather than working with these strong forms directly, the hybrid formulation casts both equations simultaneously into a single weak statement, which is the standard route for avoiding the spurious modes that plague nodal-only discretizations of the vector wave equation~\cite{Lee94}.

The weak form is obtained by multiplying each equation by test functions and integrating over the transverse cross-section $\Omega$. Let $\mathbf{v}_t \in H(\mathrm{curl};\Omega)$ be a vector test function associated with $\mathbf{E}_t$ and $v_z \in H^1(\Omega)$ a scalar test function associated with $E_z$. Multiplying \eqref{eq:transverse} by $\mathbf{v}_t$,
multiplying \eqref{eq:longitudinal} by $v_z$, integrating over $\Omega$, applying integration by parts to transfer one curl or gradient from the trial function to the test function, and discarding boundary terms (which vanish under either PEC or open boundary conditions), one arrives at the following bilinear forms. Defining the combined trial field $\mathbf{E} = (\mathbf{E}_t, E_z)$ and combined test field $\mathbf{v} = (\mathbf{v}_t, v_z)$,
\begin{equation}
A(\mathbf{E},\mathbf{v}) =
\int_\Omega \left[
\frac{1}{\mu_r} \frac{(\nabla_t \times \mathbf{E}_t)(\nabla_t \times \mathbf{v}_t)}{k_0^2}
- \epsilon_r \, \mathbf{E}_t \cdot \mathbf{v}_t
+ \frac{1}{\mu_r} (\nabla_t E_z) \cdot \mathbf{v}_t
+ \epsilon_r \, \mathbf{E}_t \cdot \nabla_t v_z
- \epsilon_r k_0^2 E_z v_z
\right] dA,
\label{eq:bilinA}
\end{equation}
\begin{equation}
B(\mathbf{E},\mathbf{v}) =
- \int_\Omega \frac{1}{\mu_r} \frac{\mathbf{E}_t \cdot \mathbf{v}_t}{k_0^2} \, dA.
\label{eq:bilinB}
\end{equation}
The first term in Eq. \eqref{eq:bilinA} penalizes spurious rotational field components through the curl--curl stiffness. The second and fifth terms are mass contributions from the transverse and longitudinal permittivities, respectively. The third and fourth terms couple $E_z$ to $\mathbf{E}_t$ in a skew-symmetric fashion that preserves self-adjointness of the system.
The right-hand-side bilinear form \eqref{eq:bilinB} is a scaled transverse mass matrix. The eigenvalue problem is then
\begin{equation}
A(\mathbf{E},\mathbf{v}) = \lambda \, B(\mathbf{E},\mathbf{v}),
\qquad \forall\,\mathbf{v},
\label{eq:evp_weak}
\end{equation}
with eigenvalue $\lambda = \beta^2/k_0^2 = n_\mathrm{eff}^2$, so that the effective index of each guided mode is $n_\mathrm{eff} = \beta/k_0$.

The transverse domain $\Omega$ is triangulated into $N_e$ linear triangles. The cross-section consists of three distinct material regions: the waveguide core, the top cladding, and the buried-oxide (BOX) substrate. Within each triangle the relative permittivity $\epsilon_r$ is constant and equal to the value at the element centroid, which coincides with the permittivity of whichever region contains that centroid. The mesh is generated by seeding dense arrays of
points along every material interface---the core sidewalls, the core top and bottom, and the outer domain boundary---before applying a Delaunay triangulation. This seeding strategy ensures that interface-aligned edges are present without requiring a constrained triangulation algorithm.

Each triangle $\Omega_e$ has three vertices with coordinates $(x_1,y_1)$, $(x_2,y_2)$, $(x_3,y_3)$ and three edges. The affine map from the reference element with coordinates $(\xi,\eta)$ to the physical element is
\[
\begin{pmatrix} x \\ y \end{pmatrix}
= \begin{pmatrix} x_1 \\ y_1 \end{pmatrix}
+ \mathbf{J}\begin{pmatrix} \xi \\ \eta \end{pmatrix},
\qquad
\mathbf{J} = \begin{pmatrix} x_2-x_1 & x_3-x_1 \\ y_2-y_1 & y_3-y_1 \end{pmatrix}.
\]
The element area is $|\Omega_e| = |\det\mathbf{J}|/2$. The barycentric coordinates
$\lambda_1,\lambda_2,\lambda_3$ satisfy $\lambda_1+\lambda_2+\lambda_3=1$ and their physical
gradients are
\[
\nabla\lambda_1 = \frac{1}{2|\Omega_e|}\begin{pmatrix}y_2-y_3\\x_3-x_2\end{pmatrix},
\quad
\nabla\lambda_2 = \frac{1}{2|\Omega_e|}\begin{pmatrix}y_3-y_1\\x_1-x_3\end{pmatrix},
\quad
\nabla\lambda_3 = \frac{1}{2|\Omega_e|}\begin{pmatrix}y_1-y_2\\x_2-x_1\end{pmatrix}.
\]
These gradients are constant within each triangle, a property that simplifies both the construction of basis functions and the analytical evaluation of curl terms.

The finite element discretization uses a mixed formulation. The transverse field $\mathbf{E}_t$ is expanded in the lowest-order Nédélec edge elements \cite{Nedelec1980} and the longitudinal field $E_z$ is expanded in first-order Lagrange (nodal, P1) elements:
\[
\mathbf{E}_t \approx \sum_{i=1}^{N_\mathrm{edge}} e_i \, \mathbf{W}_i(x,y),
\qquad
E_z \approx \sum_{j=1}^{N_n} u_j \, \phi_j(x,y),
\]
where $N_\mathrm{edge}$ is the total number of mesh edges, $N_n$ is the total number of mesh nodes, $\mathbf{W}_i$ are the Nédélec vector basis functions, and $\phi_j$ are the standard hat (nodal) basis functions. The total number of degrees of freedom is $N_\mathrm{dof} = N_\mathrm{edge} + N_n$, with the global unknown vector ordered as
\[
\mathbf{x} = \begin{bmatrix} \mathbf{e} \\ \mathbf{u} \end{bmatrix} \in \mathbb{C}^{N_\mathrm{dof}},
\]
where $\mathbf{e} \in \mathbb{C}^{N_\mathrm{edge}}$ collects the edge degrees of freedom and $\mathbf{u} \in \mathbb{C}^{N_n}$ the nodal degrees of freedom.

Within element $\Omega_e$, the three local edges are labeled $k = 1,2,3$ with each edge $k$ connecting local node $i_k$ to local node $j_k$, following the pairing convention $(i_1,j_1)=(2,3)$, $(i_2,j_2)=(3,1)$, $(i_3,j_3)=(1,2)$. The local Nédélec basis function for edge $k$ is
\begin{equation}
\mathbf{W}_k = s_k \, \ell_k \left( \lambda_{i_k} \nabla\lambda_{j_k}
                                   - \lambda_{j_k} \nabla\lambda_{i_k} \right),
\label{eq:nedelec}
\end{equation}
where $\ell_k = \|\mathbf{x}_{j_k} - \mathbf{x}_{i_k}\|$ is the physical edge length and
$s_k \in \{+1,-1\}$ is an orientation sign chosen so that the global basis function has a consistent direction across elements sharing that edge. Specifically, $s_k = +1$ if the local node ordering $i_k \to j_k$ agrees with the global edge orientation (i.e.\ $\mathrm{global}(i_k) < \mathrm{global}(j_k)$) and $s_k = -1$ otherwise. The key properties of \eqref{eq:nedelec} are that the tangential component $\mathbf{W}_k \cdot \hat{t}_k$ is constant and equal to $\pm 1$ along edge $k$ and zero on the other two edges, which guarantees
tangential continuity of $\mathbf{E}_t$ across element boundaries while allowing normal discontinuities consistent with dielectric interface conditions.

The two-dimensional curl of the Nédélec basis function is constant within each element and evaluates to
\begin{equation}
\nabla_t \times \mathbf{W}_k
= s_k \, \ell_k \cdot 2\!\left(\nabla\lambda_{i_k} \times \nabla\lambda_{j_k}\right)
= s_k \, \ell_k \cdot 2\!\left[(\nabla\lambda_{i_k})_x(\nabla\lambda_{j_k})_y
                               - (\nabla\lambda_{i_k})_y(\nabla\lambda_{j_k})_x\right],
\label{eq:curlW}
\end{equation}
where the cross product of two planar vectors is understood as the scalar $z$-component of their three-dimensional cross product. Because both the Nédélec curl \eqref{eq:curlW} and the
nodal gradients $\nabla\lambda_k$ are constant over the element, only the Nédélec basis functions themselves and the P1 shape functions $\phi_k = \lambda_k$ carry spatial dependence through the barycentric coordinates at the quadrature points.

Substituting the discrete expansions into the weak form \eqref{eq:evp_weak} and choosing test functions equal to each basis function in turn yields the $6\times 6$ element matrices $\mathbf{A}_e$ and $\mathbf{B}_e$. The local degrees of freedom are ordered as $[\text{edge}_1, \text{edge}_2, \text{edge}_3, \text{node}_1, \text{node}_2, \text{node}_3]$, giving six local DOFs per element. The entries are computed by numerical integration:
\begin{equation}
(\mathbf{A}_e)_{ij} = \sum_{q=1}^{n_q} w_q \, a\!\left(\mathbf{W}^{(q)}, \phi^{(q)}, \mathbf{W}^{(q)}, \phi^{(q)}\right)\bigg|_{\mathbf{x}_q} \cdot 2|\Omega_e|,
\qquad
(\mathbf{B}_e)_{ij} = \sum_{q=1}^{n_q} w_q \, b(\cdots)\bigg|_{\mathbf{x}_q} \cdot 2|\Omega_e|,
\end{equation}
where $\{(\xi_q,\eta_q), w_q\}$ are quadrature points and weights on the reference triangle.
The solver uses a symmetric three-point Gaussian rule with points at the midpoints of the
reference triangle edges,
\[
(\xi_1,\eta_1) = \left(\tfrac{1}{6},\tfrac{1}{6}\right),\quad
(\xi_2,\eta_2) = \left(\tfrac{2}{3},\tfrac{1}{6}\right),\quad
(\xi_3,\eta_3) = \left(\tfrac{1}{6},\tfrac{2}{3}\right),\quad
w_1 = w_2 = w_3 = \tfrac{1}{6},
\]
which is exact for polynomials up to degree two. Since the Nédélec basis functions are
linear in $(\xi,\eta)$ and their curls are constant, the integrands in $\mathbf{A}_e$ and
$\mathbf{B}_e$ are at most quadratic, so this rule integrates all terms exactly.

The four blocks of the $6\times 6$ element stiffness matrix $\mathbf{A}_e$ correspond to the
four operator pairings in the bilinear form \eqref{eq:bilinA}. For local indices $i,j \le 3$
(both edge DOFs) the contribution is
\[
(\mathbf{A}_e)_{ij} \leftarrow \sum_q w_q\left[
\frac{1}{\mu_r} \frac{(\nabla_t \times \mathbf{W}_i)(\nabla_t \times \mathbf{W}_j)}{k_0^2}
- \epsilon_r\, \mathbf{W}_i \cdot \mathbf{W}_j
\right] \cdot 2|\Omega_e|.
\]
For $i \le 3$, $j > 3$ (edge test, nodal trial, with $j' = j-3$) the coupling term is
\[
(\mathbf{A}_e)_{ij} \leftarrow \sum_q w_q \left[
\frac{1}{\mu_r} (\nabla\phi_{j'}) \cdot \mathbf{W}_i
\right] \cdot 2|\Omega_e|.
\]
For $i > 3$, $j \le 3$ (nodal test, edge trial, with $i' = i-3$) the coupling term is
\[
(\mathbf{A}_e)_{ij} \leftarrow \sum_q w_q \left[
\epsilon_r\, \mathbf{W}_j \cdot \nabla\phi_{i'}
\right] \cdot 2|\Omega_e|.
\]
For $i,j > 3$ (both nodal DOFs, with $i' = i-3$, $j' = j-3$) the longitudinal mass term is
\[
(\mathbf{A}_e)_{ij} \leftarrow \sum_q w_q \left[
-\epsilon_r\, k_0^2\, \phi_{i'} \phi_{j'}
\right] \cdot 2|\Omega_e|.
\]
The element mass matrix $\mathbf{B}_e$ is nonzero only in the edge--edge block,
\[
(\mathbf{B}_e)_{ij} \leftarrow -\sum_q w_q \left[
\frac{1}{\mu_r} \frac{\mathbf{W}_i \cdot \mathbf{W}_j}{k_0^2}
\right] \cdot 2|\Omega_e|, \qquad i,j \le 3,
\]
with all nodal rows and columns equal to zero.

The global matrices $\mathbf{A}$ and $\mathbf{B}$ of size $N_\mathrm{dof} \times N_\mathrm{dof}$
are assembled from the element contributions by the standard scatter operation. Each local DOF
index is mapped to a global DOF index: edge DOFs map to their global edge number in
$\{1,\ldots,N_\mathrm{edge}\}$ and nodal DOFs map to their global node number offset by
$N_\mathrm{edge}$, i.e.\ $N_\mathrm{edge} + \text{node index}$. The global matrices are stored
in compressed sparse row (CSR) format to allow efficient matrix--vector products during the
eigensolve.

Boundary conditions are imposed by eliminating constrained degrees of freedom from the system.
For open (radiative) boundaries the longitudinal field $E_z$ is set to zero on all nodes that
lie on the outer domain boundary, which is equivalent to a first-order absorbing condition for
modes well confined to the core. These boundary nodes are identified as those appearing in
exactly one triangle, i.e.\ nodes whose associated half-edges have count one in the global
half-edge list. The constrained DOF indices are $\{N_\mathrm{edge} + n \mid n \in \mathcal{B}_n\}$,
where $\mathcal{B}_n$ is the set of boundary node indices. When metallic (PEC) walls are
instead desired, the tangential electric field must vanish on $\partial\Omega$, which requires
setting to zero all edge DOFs on boundary edges. A boundary edge is defined as one shared by
exactly one triangle, identified using the same half-edge counting procedure applied to the
edge table. After identifying the constrained DOF set $\mathcal{C}$, the free DOFs are
$\mathcal{F} = \{1,\ldots,N_\mathrm{dof}\} \setminus \mathcal{C}$, and the reduced system
\[
\mathbf{A}_\mathcal{F} \mathbf{x}_\mathcal{F} = \lambda\, \mathbf{B}_\mathcal{F} \mathbf{x}_\mathcal{F},
\quad
\mathbf{A}_\mathcal{F} = \mathbf{A}[\mathcal{F},\mathcal{F}],
\quad
\mathbf{B}_\mathcal{F} = \mathbf{B}[\mathcal{F},\mathcal{F}],
\]
is formed by extracting the rows and columns corresponding to free DOFs.

The reduced generalized eigenvalue problem is solved using the shift-invert spectral
transformation. Given a target value $\sigma \approx \lambda_\mathrm{target}$, the
shift-invert transformation converts the problem to one whose dominant eigenpairs correspond
to the eigenvalues of the original problem closest to $\sigma$. In the Python implementation, the problem is passed to the ARPACK-based implicitly restarted Arnoldi method via \texttt{scipy.sparse.linalg.eigs}, using the call
\[
\texttt{eigs}(\mathbf{A}_\mathcal{F},\; k,\; M{=}\mathbf{B}_\mathcal{F},\;
\sigma{=}\sigma).
\]
A similar expression is used in the MATLAB implementation. The
shift $\sigma$ is set automatically to $0.9\,k_0^2\,\max(\epsilon_r)$ unless an explicit
effective index guess $n_\mathrm{guess}$ is provided, in which case $\sigma = k_0^2\,n_\mathrm{guess}^2$.
The solver requests $k = N_\mathrm{modes}$ eigenpairs and uses a convergence tolerance of
$10^{-10}$ with a maximum of 500 Arnoldi iterations. The returned eigenvalues are sorted in
descending order of $\mathrm{Re}(\sqrt{\lambda})$ so that the most confined (highest-index)
modes appear first.

From each eigenvalue--eigenvector pair $(\lambda_m, \mathbf{x}_m)$ the propagation constant
and effective index are recovered as
\[
\beta_m = \sqrt{\lambda_m},
\qquad
n_{\mathrm{eff},m} = \frac{\beta_m}{k_0}.
\]
The eigenvector is expanded back to the full DOF space by inserting zeros at the constrained
indices, yielding the edge coefficient vector $\mathbf{e}_m \in \mathbb{C}^{N_\mathrm{edge}}$
and the nodal coefficient vector $\mathbf{u}_m \in \mathbb{C}^{N_n}$. The physical electric
field within element $\Omega_e$ is then reconstructed at any point $(\xi,\eta)$ as
\[
\mathbf{E}_t\big|_{\Omega_e}(\xi,\eta)
= \sum_{k=1}^{3} e_{\mathrm{edge}(e,k)} \, \mathbf{W}_k(\xi,\eta),
\qquad
E_z\big|_{\Omega_e}(\xi,\eta)
= \sum_{k=1}^{3} u_{\mathrm{node}(e,k)} \, \lambda_k(\xi,\eta),
\]
where $\mathrm{edge}(e,k)$ and $\mathrm{node}(e,k)$ are the global edge and node indices of
element $e$, respectively.

The polarization state of each mode is characterized by the TE fraction, defined as the
fraction of transverse electric energy carried by the $x$-component of the field,
\begin{equation}
f_\mathrm{TE} = \frac{\displaystyle\int_\Omega |E_x|^2 \, dA}
                     {\displaystyle\int_\Omega \left(|E_x|^2 + |E_y|^2\right) dA}.
\label{eq:te_frac}
\end{equation}
This integral is evaluated by the same three-point quadrature rule used for element assembly.
A value of $f_\mathrm{TE}$ close to unity indicates a dominant $x$-polarized (quasi-TE) mode,
while a value close to zero indicates a quasi-TM mode. The TM fraction is simply
$f_\mathrm{TM} = 1 - f_\mathrm{TE}$.

When the overlap between two modes $m$ and $n$ is needed---for instance, to verify
orthogonality or to estimate coupling coefficients in a mode expansion---the solver computes
the simplified transverse overlap integral
\begin{equation}
\mathcal{O}_{mn} = \frac{1}{2}
\int_\Omega \mathbf{E}_{t,m}^* \cdot \mathbf{E}_{t,n} \, dA,
\label{eq:overlap}
\end{equation}
again by element-wise Gaussian quadrature. For co-propagating modes in a non-magnetic medium, this approximation is proportional to the rigorous Poynting-vector overlap; a complete expression would require reconstructing $\mathbf{H}_t$ from the curl equations, which is straightforward but omitted here for brevity.

\section{Numerical Results}
\subsection*{Case Study 1}
To validate the mode solver, we consider a rectangular Si$_3$N$_4$ waveguide embedded in SiO$_2$ cladding. The waveguide width is $W = \SI{1.6}{\micro m}$, the height is $H = \SI{0.7}{\micro m}$. The total simulation window width and height are $w_\mathrm{sim} = \SI{6.0}{\micro m}$ and $h_\mathrm{sim} = \SI{4.7}{\micro m}$. The free-space operating wavelength is $\lambda = \SI{1.55}{\micro m}$. At this wavelength, the Sellmeier models (embedded in the code) yield $n_\mathrm{core} = 1.9964$ for Si$_3$N$_4$ and $n_\mathrm{clad} = 1.444$ for SiO$_2$. The mesh resolution parameter is set to $\texttt{mesh\_res} = \lfloor w_\mathrm{sim}/\lambda \times 200$, which controls the seeding density on the outer boundary.

\begin{figure}[h]
    \centering
    \includegraphics[width=0.95\linewidth]{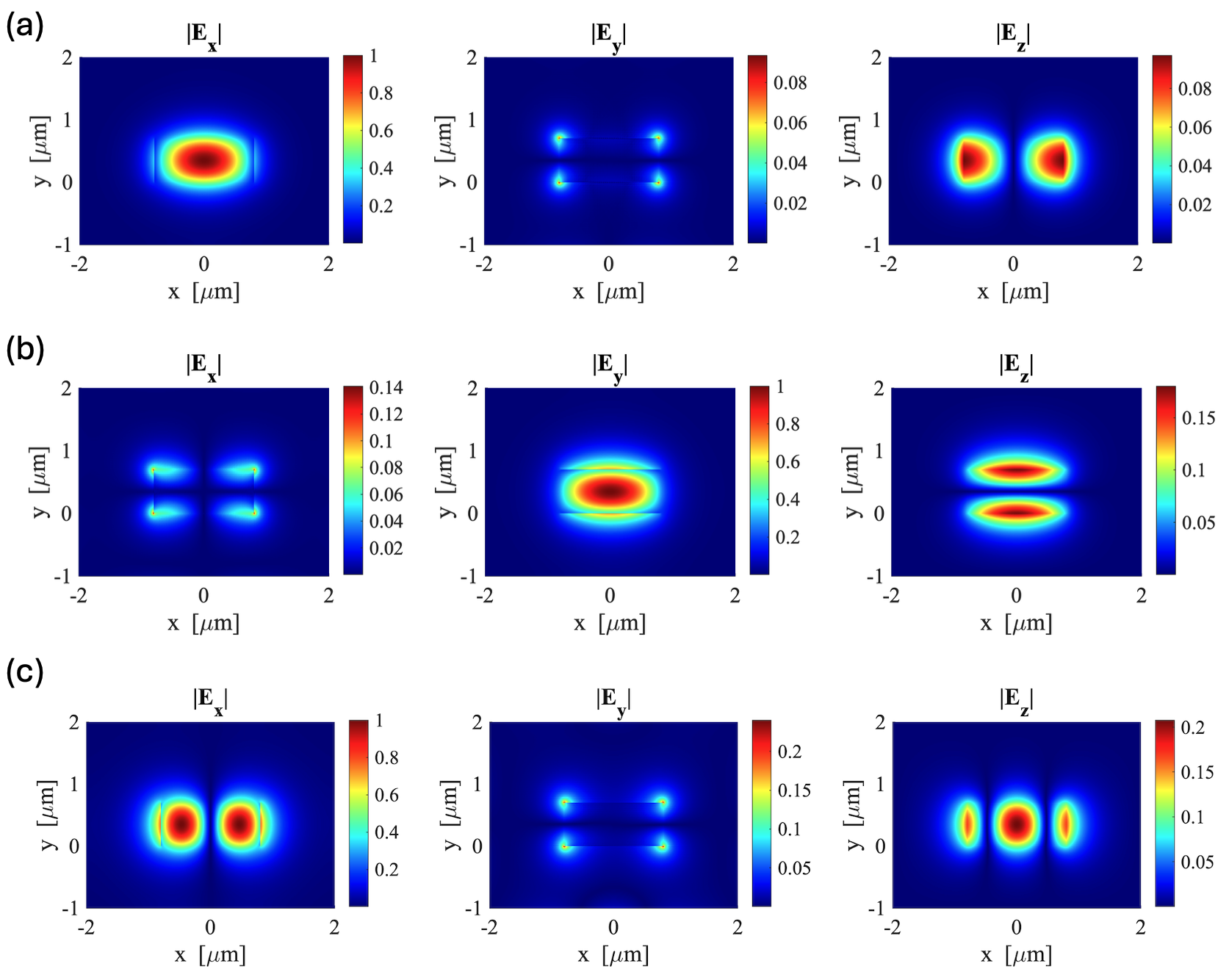}
    \caption{Normalized field components $|E_x|$, $|E_y|$, $|E_z|$ of the first three modes}
    \label{fig:mode123}
\end{figure}

Figure \ref{fig:mode123} show normalized field components $|E_x|$, $|E_y|$, $|E_z|$ of the first three modes.

Table~\ref{tab:mode_comparison} compares the effective indices obtained by the present solver against reference values from COMSOL Multiphysics. The error less than $0.004\,\%$ is observed across all four modes, confirming that the mixed Nédélec--Lagrange discretization and the shift-invert eigensolver together reproduce the guided-mode spectrum with high fidelity on an unstructured triangular mesh.

\begin{table}[h]
\centering
\caption{Comparison of effective indices determined by COMSOL Multiphysics and the present FEM solver.}
\label{tab:mode_comparison}
\begin{tabular}{
    c
    S[table-format=1.15]
    S[table-format=1.15]
    S[table-format=1.4]
}
\toprule
{Mode} & {COMSOL Multiphysics} & {Python FEM} & {Error (\%)} \\
\midrule
1 & 1.793112981078253 & 1.793062870073336 & 0.0028 \\
2 & 1.751998383202112 & 1.751994114122944 & 0.0002 \\
3 & 1.646924391702016 & 1.646884667213558 & 0.0024 \\
4 & 1.626967082269770 & 1.627030191022317 & 0.0039 \\
\bottomrule
\end{tabular}
\end{table}

\begin{figure}[h]
    \centering
    \includegraphics[width=0.45\linewidth]{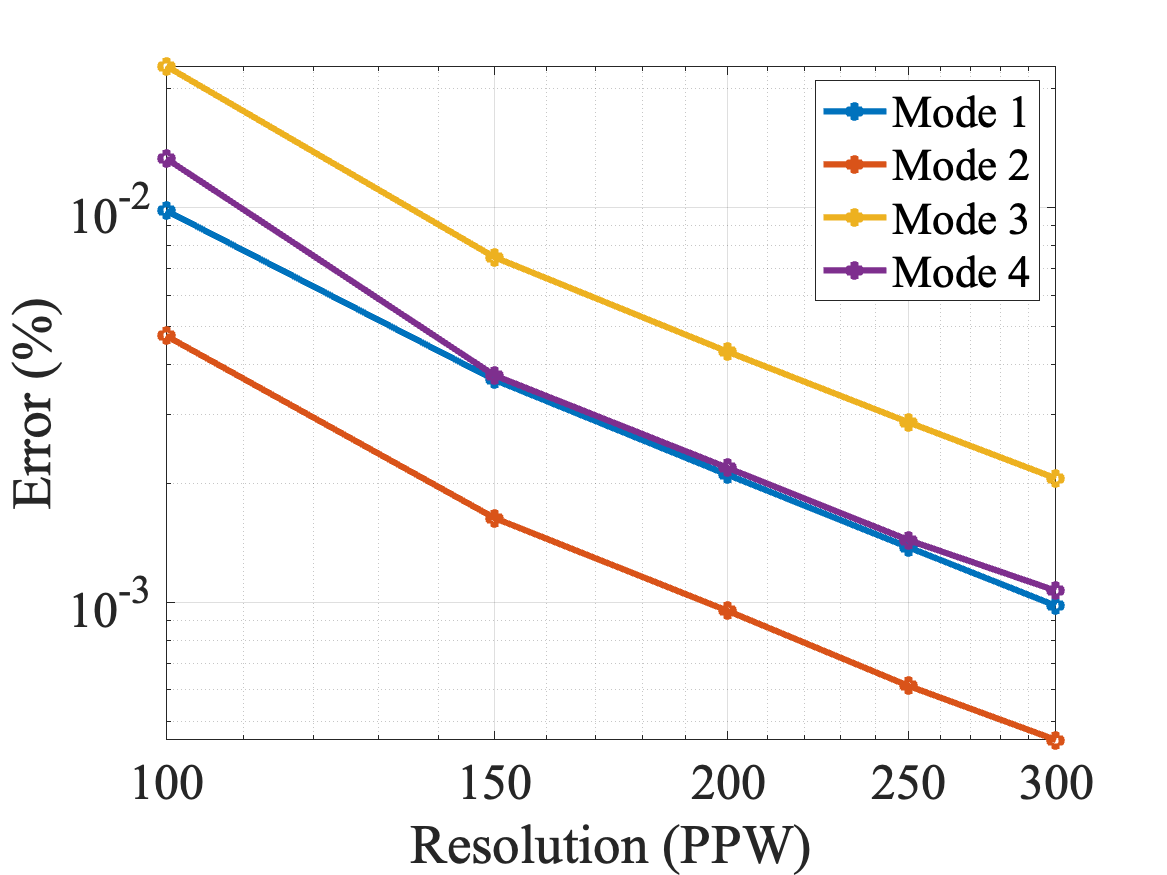} \hfill
    \includegraphics[width=0.45\linewidth]{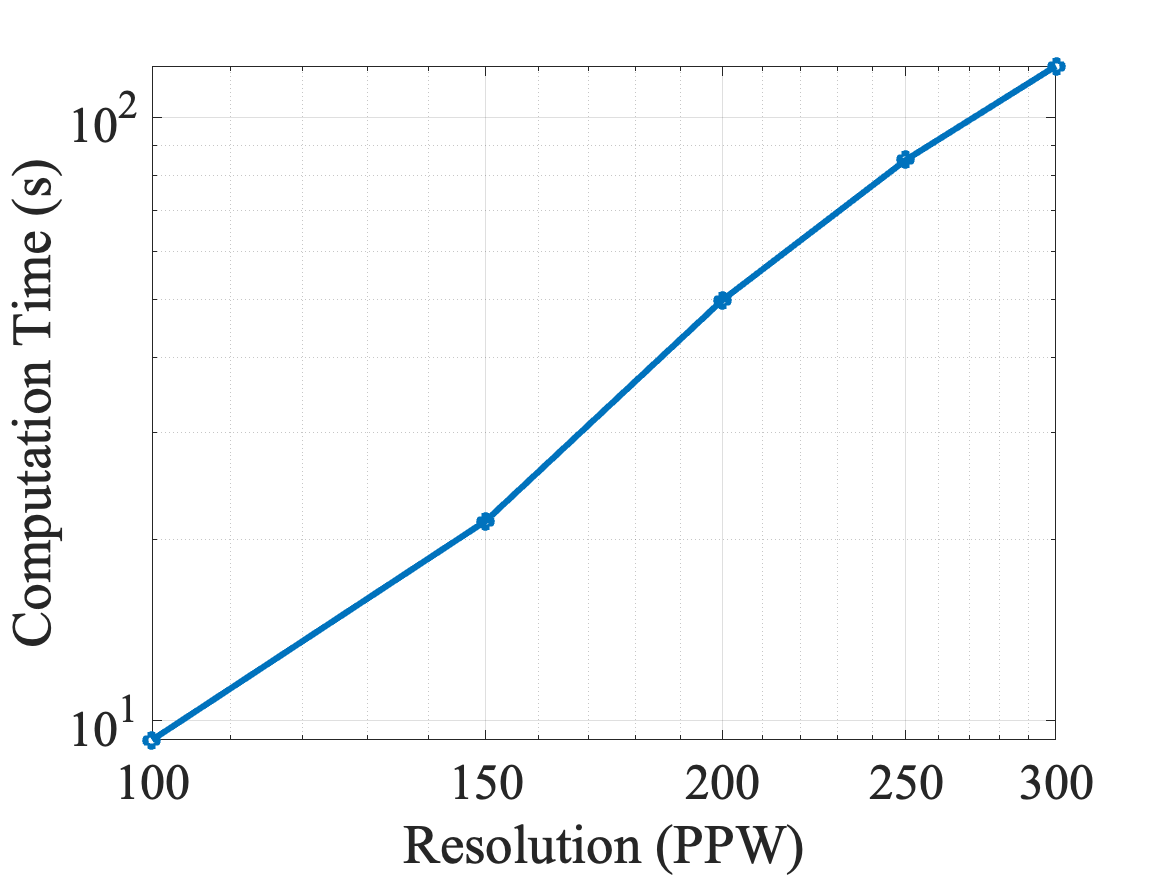}    
    \caption{Mesh quality (points per wavelength) vs. (left) error and (right) computation time. Both figures are on a log-log scale.}
    \label{fig:performance}
\end{figure}

In Fig. \ref{fig:performance}, we plot error and time as a function of mesh quality on a log-log scale. The results demonstrate a clear and consistent convergence behavior of the finite element solver as the mesh quality is improved. As the resolution, expressed in points per wavelength (PPW), increases, the relative error with respect to COMSOL Multiphysics reference solutions systematically decreases for all computed modes, confirming the accuracy and stability of the formulation. Lower-order modes exhibit smaller initial errors, while higher-order modes show greater sensitivity to coarse discretization, yet all modes converge to very low error levels at finer meshes. At the same time, the computational cost increases monotonically with mesh refinement, reflecting the expected growth in the number of degrees of freedom and the associated eigenvalue problem complexity. Although the MATLAB implementation is observed to be faster than its Python counterpart, both remain slower than COMSOL Multiphysics for very high PPW cases, which is anticipated due to COMSOL's highly optimized numerical solvers and adaptive meshing capabilities. Overall, the results validate both the correctness and robustness of the developed FEM solver while highlighting the trade-off between accuracy and computational efficiency.

\subsection*{Case Study 2}
As a second validation example, we consider a high-index-contrast rectangular dielectric waveguide with refractive index $n_\mathrm{core} = 3.5$, embedded in SiO$_2$ cladding with $n_\mathrm{clad} = 1.444$. The waveguide core has width $W = \SI{1.0}{\micro m}$ and height $H = \SI{0.6}{\micro m}$. The upper cladding thickness is $h_\mathrm{clad} = \SI{2.6}{\micro m}$, and the buried oxide (BOX) thickness is $h_\mathrm{box} = \SI{2.0}{\micro m}$. The total simulation window width and height are $w_\mathrm{sim} = \SI{6.0}{\micro m}$ and $h_\mathrm{sim} = \SI{4.6}{\micro m}$ both in our solver and COMLSOL Multiphysics. The operating wavelength is again $\lambda = \SI{1.55}{\micro m}$.

The refractive index contrast in this case is significantly higher than in Case Study~1, leading to stronger mode confinement and a larger number of guided modes. In COMSOL, the mesh is generated using an unstructured triangular discretization, with resolution controlled in the same manner as in the previous example. In MATLAB, we set the PPW to 120.

The COMSOL simulation involves $203{,}737$ degrees of freedom and completes in approximately $9$ seconds. The present MATLAB-based FEM solver uses $180{,}462$ edge-based unknowns for transverse electric field components and $60{,}701$ nodal unknowns for the longitudinal component, resulting in a total of $241{,}163$ degrees of freedom. The total computation time is approximately $9.5$ seconds.

Table \ref{tab:mode_comparison2} provides a comparison of effective indices determined by COMSOL Multiphysics and the present FEM solver. The agreement between the two solvers remains excellent, with relative errors below $0.05\,\%$ for all four modes. The results confirm that the mixed Nédélec--Lagrange formulation combined with the shift-invert eigensolver accurately captures the modal spectrum even in high-index-contrast waveguiding structures.

\begin{table}[h]
\centering
\caption{Comparison of effective indices determined by COMSOL Multiphysics and the present FEM solver for Case Study 2.}
\label{tab:mode_comparison2}
\begin{tabular}{
    c
    S[table-format=1.15]
    S[table-format=1.15]
    S[table-format=1.4]
}
\toprule
{Mode} & {COMSOL Multiphysics} & {MATLAB FEM} & {Error (\%)} \\
\midrule
1 & 3.260576219212838 & 3.260125958573306 & 0.0138 \\
2 & 3.205623278611855 & 3.205221175398172 & 0.0125 \\
3 & 2.990337795170359 & 2.989493346753727 & 0.0282 \\
4 & 2.988327221350961 & 2.987122525588838 & 0.0403 \\
\bottomrule
\end{tabular}
\end{table}

\section{Future Work}
We will focus on improving computational efficiency through advanced preconditioning techniques, parallelization strategies, and adaptive mesh refinement. We will expand our material database and add new functionalities such as a mesh generator for trapezoidal cross sections. Extensions to include anisotropic and nonlinear materials, as well as ring resonators, would further enhance our studies on integrated photonic devices and optical frequency combs.

\section{Conclusion}
In this work, we have presented a full-vector finite element mode solver for dielectric waveguides based on a mixed Nédélec--Lagrange discretization of Maxwell’s curl equations in the frequency domain. The formulation, while grounded in well-established theory, has been carefully implemented to provide a robust, efficient, and reproducible computational tool suitable for both MATLAB and Python environments. The numerical results demonstrate that the proposed solver achieves excellent agreement with reference solutions obtained from COMSOL Multiphysics, with errors on the order of $10^{-3}\,\%$ for practical mesh resolutions. The convergence study confirms the expected behavior of the finite element method, with systematic error reduction as the mesh is refined, while also highlighting the increased computational cost associated with higher resolution. The use of edge elements for transverse fields, combined with nodal elements for longitudinal components, effectively suppresses spurious modes and ensures accurate representation of hybrid electromagnetic fields. Although the computational performance of the present implementation does not yet match that of highly optimized commercial solvers, the results indicate that it provides a reliable and flexible alternative for research and educational purposes. Furthermore, the open and portable nature of the implementation enables straightforward adaptation to a wide range of waveguide configurations and material systems. 

\section*{Code Availability}
MATLAB and Python versions of the code are available at \url{https://github.com/simsekergun/Waveguide_FEM_Solver}.
The Python version can be installed and used in Jupyter Notebook or Google Colab. A sample script is provided below.

\begin{lstlisting}[language=Python]
!pip install waveguide-fem
#
from waveguide_fem import build_soi_mesh, compute_modes, get_refractive_index
from waveguide_fem import plot_mode_fields, calculate_overlap
import time
import numpy as np
from scipy.sparse import csr_matrix
from scipy.sparse.linalg import eigs
import matplotlib.pyplot as plt
import matplotlib.tri as mtri
from matplotlib.colors import Normalize
from matplotlib.cm import ScalarMappable
#
wavelength = 1.55   # microns
# Geometry (microns)
w_core = 1.6
h_core = 0.7
h_clad = 2.7
h_box  = 2.00
w_sim  = 6.00
# Materials
n_core = get_refractive_index('Si3N4', wavelength)
n_clad = get_refractive_index('SiO2',  wavelength)
n_box  = n_clad
# solver settings
num_modes        = 6    # (number of modes to search for)
ppw = 100               # mesh quality (points per wavelength)
# end of inputs
#
mesh_res         = round(w_sim / wavelength * ppw)
print(f"n_core (Si3N4) = {n_core:.6f}")
print(f"n_clad (SiO2)  = {n_clad:.6f}")
#
nodes, elems, epsilon_r, regions = build_soi_mesh(
    w_core, h_core, h_clad, h_box, w_sim,
    n_core, n_clad, n_box, mesh_res)

modes = compute_modes(nodes, elems, epsilon_r, wavelength,
                      num_modes=num_modes, mu_r=1.0)
#
print("\n--- Guided modes ---")
for m_idx, mode in enumerate(modes):
    print(f"Mode {m_idx + 1}:  n_eff = {np.real(mode['n_eff']):.6f} + "
          f"{np.imag(mode['n_eff']):.2e}i,  TE-frac = {mode['te_fraction']:.3f}")
#
# ---- Plot the dominant mode ----
plot_mode_fields(modes[0], nodes, elems, 'Mode 1 (fundamental TE)')
\end{lstlisting}

\bibliographystyle{ieeetr}
\bibliography{references}

@article{Nedelec1980,
  title={Mixed finite elements in $\cal{R}$3},
  author={Nedelec, Jean-Claude},
  journal={Numerische Mathematik},
  volume={35},
  number={3},
  pages={315--341},
  year={1980},
  publisher={Springer}
}

@ARTICLE{Wu1985,
  author={Ruey-Beei Wu and Chun Hsiung Chen},
  journal={IEEE Transactions on Microwave Theory and Techniques}, 
  title={On the Variational Reaction Theory for Dielectric Waveguides}, 
  year={1985},
  volume={33},
  number={6},
  pages={477-483},
  doi={10.1109/TMTT.1985.1133102}}

@article{koshiba1987,
  title={Finite--element method analysis of microwave and optical waveguides—trends in countermeasures to spurious solutions},
  author={Koshiba, Masanori and Hayata, Kazuya and Suzuki, Michio},
  journal={Electronics and Communications in Japan (Part II: Electronics)},
  volume={70},
  number={9},
  pages={96--108},
  year={1987},
  publisher={Wiley Online Library}
}

@article{Angkaew1987,
  title={Finite-element analysis of waveguide modes: A novel approach that eliminates spurious modes},
  author={Angkaew, Tuptim and Matsuhara, Masanori and Kumagai, Nobuaki},
  journal={IEEE transactions on microwave theory and techniques},
  volume={35},
  number={2},
  pages={117--123},
  year={1987},
  publisher={IEEE}
}

@ARTICLE{Koshiba92,
  author={Koshiba, M. and Inoue, K.},
  journal={IEEE Transactions on Microwave Theory and Techniques}, 
  title={Simple and efficient finite-element analysis of microwave and optical waveguides}, 
  year={1992},
  volume={40},
  number={2},
  pages={371-377},
  doi={10.1109/22.120111}}

@ARTICLE{Lee94,
  author={Lee, J.-F.},
  journal={IEEE Transactions on Microwave Theory and Techniques}, 
  title={Finite element analysis of lossy dielectric waveguides}, 
  year={1994},
  volume={42},
  number={6},
  pages={1025-1031},
  doi={10.1109/22.293572}}

@article{Nuno1997,
  title={Analysis of general lossy inhomogeneous and anisotropic waveguides by the finite-element method (FEM) using edge elements},
  author={Nuno, Luis and Balbastre, Juan V and Castane, Hector},
  journal={IEEE transactions on microwave theory and techniques},
  volume={45},
  number={3},
  pages={446--449},
  year={1997},
  publisher={IEEE}
}

@article{Selleri2001,
	author = {Selleri, S. and Vincetti, L. and Cucinotta, A. and Zoboli, M.},
	date = {2001/04/01},
	doi = {10.1023/A:1010886632146},
	id = {Selleri2001},
	isbn = {1572-817X},
	journal = {Optical and Quantum Electronics},
	number = {4},
	pages = {359--371},
	title = {Complex FEM modal solver of optical waveguides with PML boundary conditions},
	url = {https://doi.org/10.1023/A:1010886632146},
	volume = {33},
	year = {2001}}

@article{SelleriPetracek2001,
  author    = {Selleri, S. and Petracek, J.},
  title     = {Modal analysis of rib waveguide through finite element method},
  journal   = {Optical and Quantum Electronics},
  year      = {2001},
  volume    = {33},
  number    = {4/5},
  pages     = {373--386},
  doi = {10.1023/A:1010838716217}
}

@article{Rahman2002,
  title={Analysis of optical waveguide discontinuities},
  author={Rahman, BM Azizur and Davies, J Brian},
  journal={Journal of Lightwave Technology},
  volume={6},
  number={1},
  pages={52--57},
  year={2002},
  publisher={IEEE}
}

@article{VD2003,
  author    = {Vardapetyan, L. and Demkowicz, L.},
  title     = {Full-wave analysis of dielectric waveguides at a given frequency},
  journal   = {Mathematics of Computation},
  volume    = {72},
  number    = {241},
  pages     = {105--129},
  year      = {2003},
  publisher = {American Mathematical Society},
  doi       = {10.1090/S0025-5718-02-01411-4},
  url       = {https://doi.org/10.1090/S0025-5718-02-01411-4}
}

@ARTICLE{Murphy2008,
  author={Fallahkhair, Arman B. and Li, Kai S. and Murphy, Thomas E.},
  journal={Journal of Lightwave Technology}, 
  title={Vector Finite Difference Modesolver for Anisotropic Dielectric Waveguides}, 
  year={2008},
  volume={26},
  number={11},
  pages={1423-1431},
  doi={10.1109/JLT.2008.923643}}

@article{Simsek2025,
  author    = {Simsek, E.},
  title     = {Practical Vectorial Mode Solver for Dielectric Waveguides Based on Finite Differences},
  journal   = {Optics Letters},
  year      = {2025},
  volume    = {50},
  number    = {12},
  pages     = {4102--4105},
  doi = {10.1364/OL.550820}
}

\end{document}